\documentclass[aps,prd,amssymb,eqsecnum,showpacs,nofootinbib,twocolumn]{revtex4-1}

\usepackage{amsmath}
\usepackage{graphics}

\allowdisplaybreaks

\newcommand{\be}{\begin{equation}}

\newcommand{\ee}{\end{equation}}

\newcommand{\md}{\mathrm{d}}

\newcommand{\pp}{(p^2)}
\newcommand{\ppn}{p^2}
\newcommand{\np}{(np)}

\newcommand{\bea}{\begin{eqnarray}}
\newcommand{\eea}{\end{eqnarray}}
\newcommand{\ben}{\begin{eqnarray*}}
\newcommand{\een}{\end{eqnarray*}}

\newcommand{\myinta}{\int\!\!d^3\!x\,}

\def\pitt#1#2{\pi^{#1#2}_{\rm TT}}
\def\pittv#1#2{\pi^{#1#2}_{\rm TT}}

\def\htt#1#2{h^{\rm TT}_{#1#2}}
\def\httiv#1#2{h^{\rm TT}_{#1#2}}

\def\httivdot#1#2{\dot{h}^{\rm TT}_{#1#2}}

\begin{document}

\title{
Towards the 4th post-Newtonian Hamiltonian for two-point-mass systems}

\author{Piotr Jaranowski}
\email{pio@alpha.uwb.edu.pl}
\affiliation{Faculty of Physics,
University of Bia{\l}ystok,
Lipowa 41, 15--424 Bia{\l}ystok, Poland}

\author{Gerhard Sch\"afer}
\email{gos@tpi.uni-jena.de}
\affiliation{Theoretisch-Physikalisches Institut,
Friedrich-Schiller-Universit\"at,
Max-Wien-Pl.\ 1, 07743 Jena, Germany}

\date{}

\begin{abstract}

The article presents the conservative dynamics of gravitationally interacting two-point-mass systems    
up to the eight order in the inverse power of the velocity of light,
i.e.\ 4th post-Newtonian (4PN) order, and up to quadratic order in Newton's gravitational constant.
Additionally, all logarithmic terms at the 4PN order are given
as well as terms describing the test-mass limit.
With the aid of the Poincar\'e algebra additional terms are obtained.
The dynamics is presented in form of an autonomous Hamiltonian derived 
within the formalism of Arnowitt, Deser and Misner.
Out of the 57 different terms of the 4PN Hamiltonian in the center-of-mass frame,
the coefficients of 45 of them are derived.
Reduction of the obtained results to circular orbits is performed resulting in the 4PN-accurate formula
for energy expressed in terms of angular frequency in which two coefficients are obtained for the first time.

%\vspace{0.5cm}\noindent PACS number(s): 04.25.Nx, 04.30.Db, 97.60.Jd, 97.60.Lf

\end{abstract}


\maketitle

\section{Introduction}

Calculations of post-Newtonian (PN) expressions within general relativity have proved very
useful because of their full-fledged analytic structures through very high orders in the approximation scheme,
wherein the $n$th PN order denotes the order $(1/c^2)^n$ with $c$ denoting the speed of light.  
For compact binaries without spinning components the dynamics is explicitly known through 
3.5PN order whereby the orders 2.5PN and 3.5PN are of dissipative type, see e.g.\ \cite{JS97,DJS01,I04,BDEF04,NB05,I09,FS11a}.
Partial knowledge is available at higher orders. 
To all orders in $1/c$, the $n$-body dynamics is known through linear order in Newton's gravitational constant $G$ \cite{LSB08},
and in the order $1/c^8$ the terms quadratic in $G$ were presented quite recently \cite{FS12}. 
The terms at the 4PN level containing logarithms are known too \cite{BD88,D10,BDTW10}. 

In case of spinning binaries the conservative dynamics is known through 3PN order in both spin-orbit and spin1-spin2 
couplings or, if the spins are counted $1/c$, 3.5PN and 4PN order, respectively, see e.g.\ \cite{HS11a,HS11b,L12}.
For the dissipative spin-orbit and spin1-spin2 couplings the highest achieved order is 3.5PN
(counting spin as of order zero) \cite{ZW07,WW07,SW10, WSZS11}.

Those calculations are very important for future gravitational wave astronomy because they allow the determination of the 
binary insipral process with high precision, \cite{LIGO11}. Even last stable binary orbits can be deduced from
them with high confidence though they are located in the limiting regime of PN approximations. 
On the other side, the application of PN results in the effective one-body approach (EOB)
even allows the transition beyond the last stable orbit regime, see e.g.\ \cite{BDGNR11,BBLT12}. 
No doubt, the knowledge of any part of the 4PN point-mass dynamics is of great importance. 

We employ the following notations:
$\mathbf{x}=\left(x^i\right)$, $i=1,2,3$,
denotes a point in the 3-dimensional Euclidean space $\mathbb{R}^3$
endowed with a standard Euclidean metric and a scalar product (denoted by a dot).
Letters $a,b$ are body labels ($a,b=1,2$),
so ${\bf x}_a\in\mathbb{R}^3$ denotes the position of the $a$th point mass.
We also define ${\bf r}_{12}\equiv{\bf x}_2 - {\bf x}_1$,
$r_{12}\equiv|{\bf r}_{12}|$, ${\bf n}\equiv{\bf r}_{12}/r_{12}$;
$|\cdot|$ stands here for the Euclidean length of a vector.
The linear momentum vector of the $a$th body is denoted by ${\bf p}_a=\left(p_{ai}\right)$,
and $m_a$ denotes its mass parameter.
An overdot means the total time derivative.
We abbreviate $\delta\left({\bf x}-{\bf x}_a\right)$ by $\delta_a$.
Throughout the paper we extensively used the computer-algebra system \textsc{Mathematica}.

\section{The ADM canonical approach}

In this Section we use units in which $16\pi G=c=1$.  
In the Arnowitt, Deser, and Misner (ADM) canonical approach to general relativity \cite{ADM62},
the constraint equations are the crucial equations to be solved for the description of the dynamics.
Written for many-point-mass systems they read
\begin{subequations}
\label{coneqs}
\begin{align}
g^{-1/2}\Big(gR+\frac{1}{2}\left(g_{ij}\pi^{ij}\right)^2
&- \pi_{ij}\pi^{ij}\Big)
\nonumber\\
&= \sum_a\left(\gamma^{ij}p_{ai}p_{aj}+m_a^2\right)^{1/2}\delta_a,
\\
-2{\pi^{ij}}_{|j} &= \sum_a \gamma^{ij}p_{aj}\delta_a.
\end{align}
\end{subequations}
Here, the 3-metric reads $g_{ij}$ and its inverse is denoted by $\gamma^{ij}$,
$g$ is the determinant of the 3-metric,
$R$ is the curvature scalar of the time-equal-constant slices,
and $\pi^{ij}$ the canonical conjugate to $g_{ij}$.
The 3-dimensional covariant derivative is denoted by $_|$.   
We have employed the ADMTT coordinate conditions, 
\be
\label{gauge}
g_{ij} = \left(1+\frac{1}{8}\phi\right)^4\delta_{ij}+{\htt ij},
\qquad \pi^{ii}=0,
\ee
where $h^{\rm TT}_{ij}$ is a transverse-traceless (TT) quantity.
The field momentum $\pi^{ij}$ is splitted into its longitudinal $\tilde{\pi}_{ij}$ and TT ${\pitt ij}$ parts,
$\pi^{ij}=\tilde{\pi}_{ij}+{\pitt ij}$.
By solving the constraint equations \eqref{coneqs},
the ADM Hamiltonian can be put into its reduced form,
\be
H_{\rm red}\big[{\bf x}_a,{\bf p}_{a},{\htt ij},{\pitt ij}\big]
= -\myinta \Delta\phi.
\ee
This Hamiltonian describes the evolution of the matter $({\bf x}_a,{\bf p}_{a})$
and independent gravitational field $({\htt ij},{\pitt ij})$ variables.
For the latter, $\pi^{ij}_{\rm TT}$ is the canonical conjugate to $h^{\rm TT}_{ij}$.

An autonomous conservative Hamiltonian can be obtained through the transition to a Routhian description,
\be
R\left[{\bf x}_a,{\bf p}_{a},{\httiv ij},{\httivdot ij}\right]
= H_{\rm red}  - \myinta{\pittv ij}{\httivdot ij}.
\ee
Then the matter Hamiltonian reads
\begin{align}
H({\bf x}_a,{\bf p}_{a})
= R\left[{\bf x}_a,{\bf p}_{a},{\httiv ij}({\bf x}_a,{\bf p}_{a}),{\httivdot ij}({\bf x}_a,{\bf p}_{a})\right],
\end{align}
where all time derivatives of ${\bf x}_a$ and ${\bf p}_a$ are eliminated
through lower-order Hamilton equations of motion \cite{DS91},
what is equivalent to performing a canonical transformation.
For computation of the 4PN matter Hamiltonian one needs to use Newtonian and 1PN equations of motion.

\section{Results}

We have iteratively solved, by a PN expansion up to the 4PN order,
the constraint equations \eqref{coneqs} for the functions $\phi$ and $\tilde{\pi}_{ij}$
(the iterative solution of the constraints up to the 3PN order is given in detail in Ref.\ \cite{JS98}).
We have thus obtained the 4PN-accurate Hamiltonian density.
The 3-dimensional integral over it we have regularized
by means of procedures described in Appendix B of \cite{JS98}.
To diminish number of terms we display here
the Hamiltonian at the 4PN order only and in the center-of-mass frame defined by the condition ${\bf p}_1+{\bf p}_2=0$.
We introduce the following reduced variables:
${\bf r}\equiv{\bf r}_{12}/(GM)$ (with $r\equiv|\mathbf{r}|$ and $\mathbf{n}\equiv\mathbf{r}/r$),
${\bf p}\equiv{\bf p}_1/\mu$, where $M \equiv m_1 + m_2$ is the total mass of the system
and $\mu \equiv m_1m_2/M$ is its reduced mass.
We also introduce the reduced Hamiltonian $\hat{H}\equiv(H-Mc^2)/\mu$
which depends on masses only through the symmetric mas ratio $\nu\equiv\mu/M$
($0\le\nu\le1/4$; $\nu=0$ is the test-mass limit and $\nu = 1/4$ holds for equal masses).
The 4PN reduced Hamiltonian reads
[here $\pp\equiv\mathbf{p}\cdot\mathbf{p}$ and $\np\equiv\mathbf{n}\cdot\mathbf{p}$]
\begin{widetext}
\begin{align}
\label{hatH}
c^8\,{\hat H}_\text{4PN}(\mathbf{r},\mathbf{p}) &=
\left(
\frac{7}{256}
-\frac{63}{256}\nu
+\frac{189}{256}\nu^2
-\frac{105}{128}\nu^3
+\frac{63}{256}\nu^4
\right)\pp^5
\nonumber\\
&\quad
+ \Bigg\{
\frac{45}{128}\pp^4
-\frac{45}{16}\pp^4\nu
+\left(
\frac{423}{64}\pp^4
-\frac{3}{32}\np^2\pp^3
-\frac{9}{64}\np^4\pp^2
\right)\nu^2
\nonumber\\
&\qquad\quad
+ \left(
-\frac{1013}{256}\pp^4
+\frac{23}{64}\np^2\pp^3
+\frac{69}{128}\np^4\pp^2
-\frac{5}{64}\np^6\ppn
+\frac{35}{256}\np^8
\right)\nu^3
\nonumber\\
&\qquad\quad
+ \left(
-\frac{35}{128}\pp^4
-\frac{5}{32}\np^2\pp^3
-\frac{9}{64}\np^4\pp^2
-\frac{5}{32}\np^6\ppn
-\frac{35}{128}\np^8
\right)\nu^4
\Bigg\}\frac{1}{r}
\nonumber\\
&\quad
+ \Bigg\{
\frac{13}{8}\pp^3
+ \left(
-\frac{791}{64}\pp^3
+\frac{49}{16}\np^2\pp^2
-\frac{889}{192}\np^4\ppn
+\frac{369}{160}\np^6
\right)\nu
\nonumber\\
&\qquad\quad
+ \left(
\frac{4857}{256}\pp^3
-\frac{545}{64}\np^2\pp^2
+\frac{9475}{768}\np^4\ppn
-\frac{1151}{128}\np^6
\right)\nu^2
\nonumber\\
&\qquad\quad
+ \left(
\frac{2335}{256}\pp^3
+\frac{1135}{256}\np^2\pp^2
-\frac{1649}{768}\np^4\ppn
+\frac{10353}{1280}\np^6
\right)\nu^3
\Bigg\}\frac{1}{r^2}
\nonumber\\
&\quad
+ \Bigg\{
\frac{105}{32}\pp^2
+ \left[
C_{41}(\mathbf{r},\mathbf{p})
+ \left(
\frac{237}{40}\pp^2
-\frac{1293}{40}\np^2\ppn
+\frac{97}{4}\np^4
\right) \ln\frac{r}{\hat{s}} \right]\nu
\nonumber\\
&\qquad\quad
+ C_{42}(\mathbf{r},\mathbf{p})\,\nu^2
+ \left(
-\frac{553}{128}\pp^2
-\frac{225}{64}\np^2\ppn
-\frac{381}{128}\np^4
\right)\nu^3 
\Bigg\}\frac{1}{r^3}
\nonumber\\
&\quad
+ \Bigg\{
\frac{105}{32}\ppn
+ \left[C_{21}(\mathbf{r},\mathbf{p})
+ \left(
\frac{233}{40}\ppn
-\frac{29}{6} \np^2
\right)\ln\frac{r}{\hat{s}} \right]\nu
+ C_{22}(\mathbf{r},\mathbf{p})\,\nu^2
\Bigg\}\frac{1}{r^4}
\nonumber\\
&\quad
+ \Bigg\{
-\frac{1}{16}
+ \left[c_{01}+\frac{21}{20}\ln\frac{r}{\hat{s}}\right]\nu
+ c_{02}\,\nu^2
\Bigg\}\frac{1}{r^5},
\end{align}
wherein the non computed terms $C_{4i}(\mathbf{r},\mathbf{p})$ and $C_{2i}(\mathbf{r},\mathbf{p})$ have the structure
\begin{subequations}
\begin{align}
C_{4i}(\mathbf{r},\mathbf{p}) &= c_{4i1}\pp^2 + c_{4i2}\np^2\ppn + c_{4i3}\np^4, \quad i = 1,2,
\\[1ex]
C_{2i}(\mathbf{r},\mathbf{p}) &= c_{2i1}\pp + c_{2i2}\np^2, \quad i = 1,2,
\end{align}
\end{subequations}
\end{widetext}
where $c_{4i1}$, $c_{4i2}$, $c_{4i2}$, $c_{4i3}$, $c_{2i2}$ ($i=1,2$)
together with $c_{01}$ and $c_{02}$ are not yet computed numerical coefficients.
The Hamiltonian \eqref{hatH} is composed of 57 terms, coefficients of 45 of them are computed.
All terms of the order 10, 8, and 6 in momenta are calculated.
Regularization of integrals related with these terms does not lead to any ambiguity.
We have checked that the terms of the order 8 in momenta
(they are linear in Newton's gravitational constant $G$)
are, up to adding a total time derivative,
compatible with the 4PN Hamiltonian which follow from the exact post-Minkowskian Hamiltonian derived in Ref.\ \cite{LSB08}.
We have computed all terms for $\nu=0$ and have checked that they agree with the 4PN approximation of the Hamiltonian
of a test body orbiting around a Schwarzschild black hole [see, e.g., Eq.\ (77) in \cite{JS98}].
Linear and quadratic in Newton's gravitational constant $G$ terms in Lagrangian form were recently
derived in \cite{FS12}.

We have computed all the logarithmic terms of the Hamiltonian resulting from divergences of the instantaneous 
near-zone metric when going to large distances. In these logarithmic terms $\hat{s}$ is a non-fixed constant regularization scale
[more precisley $\hat{s}$ is the reduced regularization scale, $\hat{s}\equiv{s/(GM)}$,
where $s$ is a regularization scale with dimension of length; let us note that $r_{12}/s=r/\hat{s}$].
Although the logarithms do appear in terms that we are not able to regularize non-ambiguously by means of 3-dimensional procedures,
these procedures give unique values of the coefficients of the logarithms.
The logarithms derived by us are in agreement with the results obtained earlier in \cite{D10,BDTW10}
in the sense that by means of them we have recovered a computed earlier coefficient of logarithm
in the formula for energy as a function of angular frequency along circular orbit [see Eq.\ \eqref{eofx} below]. 
Instantaneous 4PN logarithmic terms were also derived in Ref.\ \cite{FS11b} from manifest non-local-in-time terms.

We have computed but not displayed here the 4PN Hamiltonian in the general reference frame
(which depends on some more undetermined coefficients).
We have checked that this Hamiltonian respects Poincar\'e invariance in a way described in detail in Ref.\ \cite{DJS2000}.
Namely, we have constructed the most general template for the 4PN center-of-mass vector
(with all but one coefficients unspecified and with terms proportional to $\ln(r_{12}/s)$ present)
and we have checked that it is possible to fit the coefficients of this vector in such a way
that all Poincar\'e algebra relations are fulfilled with 4PN accuracy.
As a ``by-product'' of this checking we have uniquely
determined the quartic in momenta and proportional to $\nu^3$ coefficient.

The numerical coefficients not yet computed in the 4PN center-of-mass Hamiltonian \eqref{hatH},
12 different ones in total, need more complicated regularization 
procedures, likely even beyond pure dimensional regularization as successfully done e.g.\ in \cite{DJS01}.

Making use of the Hamiltonian \eqref{hatH} we have computed energy of binary system
as a function of angular frequency along circular orbits
(details of analogous computations performed at the 3PN order are given in Ref.\ \cite{DJS2000b}).
Along circular orbit $\mathbf{n}\cdot\mathbf{p}=0$ and $(\mathbf{p})^2=j^2/r^2$,
where $\mathbf{j}=\mathbf{r}\times\mathbf{p}$ (with $j\equiv|\mathbf{j}|$)
is the reduced total angular momentum of the binary system.
From equation $\partial{\hat{H}(r,j^2/r^2)}/\partial r=0$ one gets the relation $r=r(j)$
which, after substiution back to the Hamiltonian, defines energy $E$ along circular orbits as a function of $j$, $E=E(j)$.
Then the angular frequency $\omega$ of circular orbits can be computed from the relation
$\omega=(GM)^{-1}\md E/\md j$. This relation can be perturbatively inverted to give $j$ as a function of $\omega$
and finally $E=E(\omega)$. Instead of $\omega$ one often uses the dimensionless PN parameter
$x\equiv(GM\omega/c^3)^{2/3}$.
We have also assumed that the regularization scale $s$ can be related
with the period $P$ along circular orbits, $s=cP$ \cite{BDTW10}.
Making then use of the third Kepler's law at the Newtonian level
one shows that $\hat{s}=2\pi c^{-2}x^{-3/2}$.
Applying the above described procedure to the Hamiltonian
$\hat{H}_\text{N}+\hat{H}_\text{1PN}+\hat{H}_\text{2PN}+\hat{H}_\text{3PN}+\hat{H}_\text{4PN}$
[the reduced Hamiltonians from $\hat{H}_\text{N}$ to $\hat{H}_\text{3PN}$
can be found in Eq.\ (3.6) of \cite{DJS2000b},
where one has to put $\omega_{\text{static}}=0$ and $\omega_{\text{kinetic}}=41/24$,
see \cite{DJS01}],
the 4PN-accurate binding energy of the system can be put in the form
\begin{widetext}
\begin{align}
\label{eofx}
E(x;\nu) = -\frac{\mu c^2 x}{2} \biggl(
1 &+  e_{\text{1PN}}(\nu)\,x + e_{\text{2PN}}(\nu)\,x^2
+ e_{\text{3PN}}(\nu)\,x^3
\nonumber\\[0.5ex]
&+ \Big(e_{\text{4PN}}(\nu)+\frac{448}{15}\nu\ln x\Big)\,x^4
+ {\cal O}\big((v/c)^{10}\big) \biggr),
\end{align}
where the fractional corrections to the Newtonian energy at different PN orders read
\begin{subequations}
\begin{align}
e_{\text{1PN}}(\nu) &= -\frac{3}{4} - \frac{1}{12}\nu,
\\[1ex]
e_{\text{2PN}}(\nu) &= -\frac{27}{8} + \frac{19}{8}\nu
- \frac{1}{24}\nu^2,
\\[1ex]
e_{\text{3PN}}(\nu) &= -\frac{675}{64} +
\left(\frac{34445}{576}-\frac{205}{96}\pi^2\right)\nu -
\frac{155}{96}\nu^2 - \frac{35}{5184}\nu^3,
\\[1ex]
e_{\text{4PN}}(\nu) &= -\frac{3969}{128}
+ c_1\,\nu
+ c_2\,\nu^2
+ \frac{301}{1728}\nu^3
+ \frac{77}{31104}\nu^4.
\end{align}
\end{subequations}
\end{widetext}
In the above formula the two 4PN coefficients, at $\nu^3$ and $\nu^4$,
were computed for the first time. 
The 4PN coefficient $c_1$ linear in $\nu$ was computed numerically in Ref.\ \cite{TBW2012}.
Its approximate value equals [see also Eq.\ (3.1) in \cite{BBLT12}]
\be
c_1 \cong 153.8803.
\ee

\begin{acknowledgments}

P.J.\ gratefully acknowledges support of the Deutsche Forschungsgemeinschaft (DFG) through the Transregional 
Collaborative Research Center SFB/TR7 ``Gravitational Wave Astronomy: Methods--Sources--Observation''. 
The work of P.J.\ was also supported in part by the Polish MNiSW grant no.\ N N203 387237.

\end{acknowledgments}

\end{document}